\documentclass[preprint,12pt]{elsarticle}

\usepackage{amssymb}
\usepackage{graphicx}

\journal{Physics Letters A}

\begin{document}

\begin{frontmatter}

\title{Tempered relaxation with clustering patterns}

\author[label1]{Aleksander Stanislavsky\fnref{cor1}}
\fntext[cor1]{E-mail: alexstan@ri.kharkov.ua}
\author[label2]{Karina Weron\fnref{cor2}}
\fntext[cor2]{E-mail: Karina.Weron@pwr.wroc.pl}
\address[label1]{Institute of Radio Astronomy, Ukrainian National Academy of
Sciences,\\ 4 Chervonopraporna St., 61002 Kharkov, Ukraine}
\address[label2]{Institute of Physics,  Wroc{\l}aw University of Technology,\\
Wybrze${\dot z}$e Wyspia$\acute n$skiego 27, 50-370 Wroc{\l}aw,
Poland}

\begin{abstract}
This work is motivated by the relaxation data for materials which
exhibit a change of the relationship between the fractional
power-law exponents when different relaxation peaks in their
dielectric susceptibility are observed. Within the proposed
framework we derive a frequency-domain relaxation function fitting
the whole range of the two-power-law dielectric spectroscopy data
with independent low- and high-frequency fractional exponents
$\gamma$ and $-\alpha$, respectively. We show that this effect
results from a contribution of different processes.  For high
frequencies it is determined by random stops and movement of
relaxing components, and the low-frequency slope is caused by
clustering in their temporal changes.
\end{abstract}

\begin{keyword}
Tempered $\alpha$-stable process \sep Fractional two-power
relaxation \sep Compound subordination

\PACS 05.40.Fb \sep 77.22.Gm \sep 02.50.Ey

\end{keyword}

\end{frontmatter}

\section{Introduction}
\begin{figure}
\resizebox{1.\columnwidth}{!}{%
\centering
  \includegraphics{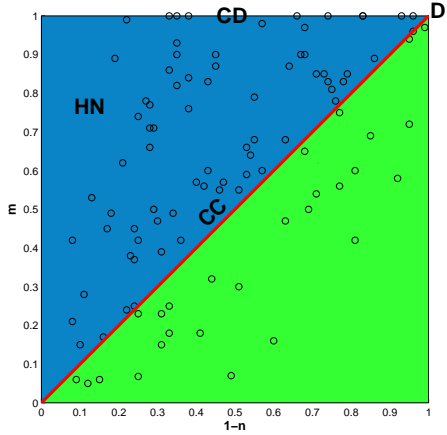}
} \caption{(Color online) Relaxation diagram positioning different
laws of relaxation. The exponents $m$ and $1-n$ focus on
declination of the imaginary susceptibility $\chi''(\omega)$ for
low and high frequencies. The circles are experimental points (for
various materials) taken from the book \cite{J2}.}\label{fig0}
\end{figure}

The stochastic approach based on subordination of compound random
processes for description of the non-exponential relaxation
phenomena within the anomalous diffusion framework shows itself as
one of the most useful mathematical tools \cite{wjmwt2010}. Its
recent success allows one to get the fractional two-power
dependency commonly observed in relaxation of complex materials
(glasses, liquid crystals, polymers, biopolymers and so on).
According to this approach, the well-known Havriliak-Negami (HN)
relaxation appears due to a compound subordination by two
independent random processes. One of them is connected  with a
coupling between the very large jumps in physical and operational
times, and another accounts for the amount of time when a relaxing
entity does not participate in motion. In fact, both of them are
described by means of a skewed $\alpha$-stable process and its
inverse. Although, from one side the results help one to
understand better the complex nature of the physical mechanisms
underlying the power-law relaxation responses, at the same time
they raise a number of problems to the theory of relaxation. In
particular, any $\alpha$-stable process has even no the first
moment. In real situations there are sufficiently many factors
truncating the distribution of $\alpha$-stable processes
\cite{Houg86,Mant94,Kop95,Boyar02,stan08}. This leads to the
tempered $\alpha$-stable processes, and they have all the moments.
As it is shown in \cite{stan09}, from the subordination by the
inverse tempered $\alpha$-stable process one can derive the
tempered diffusion equation and the relaxation function describing
the Debye (D), Cole-Cole (CC) and Cole-Davidson (CD) types of
relaxation. The tempered diffusion has a transient character, i.\
e. a crossover from subdiffusion at small times to normal
diffusion at long times. The transient subdiffusion has impact on
kinetics of magnetic bright points on the Sun \cite{cadav} and has
been observed in cells and cell membranes
\cite{platani,Wedemeier,Schmidt}. Physical arguments for
appearance of such effects are that subdiffusion is caused by
traps, but in a finite system there is a given maximal depth of
the traps (maximal waiting time) truncating their power-law
waiting time distribution in such a way that beyond the maximal
waiting time the diffusive behavior of the complex system tends to
normal. Note also that the truncation of waiting times
demonstrates features of weak ergodicity breaking in motion of
lipid granules \cite{Jeon}. The above mentioned relaxation
functions are only partial cases of the more universal HN law (see
Fig.~\ref{fig0}). Let us mention at this point that the original
HN function fits \cite{J1,J2} the fractional power-law dependence
observed in large part of the dielectric susceptibility
$\chi(\omega)$ data:
\begin{eqnarray}
\Delta\chi(\omega)=\chi(0)-\chi(\omega)&\propto&\left(i\omega/
\omega_p\right)^m\ \ \quad {\rm
for}\quad\omega\ll\omega_p\,,\label{eq0}\\
\chi(\omega)&\propto&\left(i\omega/\omega_p\right)^{n-1}\quad {\rm
for}\quad\omega\gg\omega_p\nonumber
\end{eqnarray}
with $0<m<1$ and $0<1-n<1$, the low- and high-frequency exponents,
respectively. It has to be stressed the HN exponents fulfill the
following relation $m > 1-n$. Here $\omega_p$ denotes the loss
peak frequency, $(i\omega)^\lambda$ means
$(i\omega)^\lambda=|\omega|^\lambda\,\exp(i\,\lambda\,\pi\,{\rm
sgn}(\omega)/2)$ and $i=\sqrt{-1}$. The data characterized by the
opposite relation $m<1-n$ (the diagram part of Fig.~\ref{fig0}
below the line CC) cannot be interpreted in terms of this
function, since its origins, found within the fractional
Fokker-Planck equation \cite{kcc04} and continuous time random
walk \cite{wjm05,marek} approaches, cannot yield this relation.
The corresponding time-domain description uses the notion of a
response function $f(t)=-d\phi(t)/dt$ (negative time derivative of
the relaxation function $\phi(t)$),  exhibiting the following
power-law asymptotics
\begin{displaymath}
f(t)\propto\cases{(t/\tau_p)^{-n}\,\qquad {\rm for}\quad t
\ll\tau_p\,,&\cr (t/\tau_p)^{-m-1}\quad {\rm for}\quad t
\gg\tau_p\,,&\cr}
\end{displaymath}
where $\tau_p=1/\omega_p$ is the characteristic relaxation time,
determined by the loss peak frequency. Here is useful to mention
that an ideal capacitor with an impedance of the form $1/(i\omega
C)$ does not exist in nature, and as a rule, dielectric materials
exhibit a more realistic fractional behavior $1/(i\omega
C)^\alpha$ with $\alpha<1$ \cite{boh}.

The main open question, arising here, is whether it is possible to
describe such a general relaxation picture (in the sense of
arbitrary exponents $m$ and $1-n$) by the tempered scheme. It
would be interesting and important for understanding of the
relaxation mechanism, especially when a change of the relationship
between the power exponents from $m>1-n$ to $m<1-n$ is observed in
one dielectric system for different dielectric susceptibility
peaks. As we see, the same material can have both the $m<1-n$ and
$m>1-n$ relaxation patterns under different temperature/pressure
conditions, and a model, in which this passage via model
parameters is possible, is required. It would be also of a great
importance to find a single analytical form  of the relaxation
function fulfilling such requirements (see Fig.~\ref{fig0} and
Table~\ref{tab0}).

The purpose of this Letter is to shed light on the character of
relaxation processes in such materials. To reach this goal we
present a new scenario of subordination based on finite-moment
random processes that extend the theory of tempered relaxation
\cite{stan09}.  The basic idea presented here also starts with a
compound subordination of two random processes independent of each
other. One of them is the inverse tempered $\alpha$-stable process
but the another one is the undershooting process with index
$0<\gamma<1$ \cite{wjmwt2010,stan10}.

\begin{table}
\caption{The relation between exponents $m$ and $1-n$ for  some
experimental data (see also \cite{J2}).} \label{tab0}
\begin{tabular}{cccc}
\hline\noalign{\smallskip}
Material & $m$ & $1-n$ & Comments \\
\noalign{\smallskip}\hline\noalign{\smallskip}
Polyvinylide- & 0.07 & 0.49 & $\alpha$-peak, $m<1-n$ \\
ne fluoride &  &  & \\
Polyvinylide- & 0.53 & 0.13 & $\beta$-peak, $m>1-n$ \\
ne fluoride &  &  & \\
Polyvinylide- & 0.05 & 0.12 & $\gamma$-peak, $m<1-n$ \\
ne fluoride &  &  & \\
\noalign{\smallskip}\hline
    Nylon 610 & 0.6 & 0.57 & $T =$ 483 K, $m>1-n$ \\
    Nylon 610 & 0.3 & 0.51 & $T =$ 413 K, $m<1-n$ \\
    Nylon 610 & 0.18 & 0.41 & $T =$ 373 K, $m<1-n$ \\
\noalign{\smallskip}\hline
    Glycerol & 0.55 & 0.46 & $P =$ 3.1 kb, $m>1-n$ \\
    Glycerol & 0.5 & 0.69 & $P =$ 4.4 kb, $m<1-n$ \\
\noalign{\smallskip}\hline
\end{tabular}
\end{table}

The Letter is organized as follows: at first, we briefly define the
primary random processes and formulate a modification of their
subordination representation. Next, within this approach we derive
the frequency-domain relaxation function in an analytical form. In
the last section the results of our study are discussed as applied
to relaxation phenomena.

\section{Tempering and coupling in the compound subordination}  
The tempered $\alpha$-stable process \cite{Pir05,Ros07} is
characterized by the following Laplace image of its probability
density function (pdf)
\begin{equation}
\tilde{f}_\alpha(u)=\exp\left(\delta^\alpha-(u+\delta)^\alpha\right)
\,,\label{eq1}
\end{equation}
where  the stability parameter $0<\alpha\leq1$ and the tempering
parameter $\delta\geq 0$ are constants. If $\delta$ equals to
zero, the tempered $\alpha$-stable process becomes simply
$\alpha$-stable. In other words, the parameter $\delta$ provides
just a truncation of the ordinary, long-tailed totally skewed
$\alpha$-stable distribution. The truncation leads to the random
process having all moments finite. Formula (\ref{eq1}) describes
probabilistic properties of the tempered process in terms of  the
internal (operational) time. Its inverse process may be used as a
subordinator. The pdf $g_\alpha(\tau,t)$ of the subordinator depends on
the real physical time and describes the first passage over the
temporal limit $t$. Its Laplace transform reads
\begin{eqnarray}
&&\tilde{g}_\alpha(\tau,u)=-\frac{1}{u}\frac{\partial}
{\partial\tau}\tilde{f}(u,\tau)=\nonumber\\
&&\frac{(u+\delta)^\alpha-\delta^\alpha}{u}\,
\exp\left(-\tau[(u+\delta)^\alpha-\delta^\alpha]\right)
\,.\label{eq2}
\end{eqnarray}
The inverse tempered $\alpha$-stable process accounts for motion
alternating with stops so that the temporal intervals between them
are random and with heavy tails in density. The main feature of
the process is that it occurs only for small times \cite{stan08}.

If the $\gamma$-stable process is subordinated by its inverse, the
compound subordinator behaves 1-similar (proportional to time $t$)
for long times \cite{feller}. Such a subordinator does not,
however, lead to any fractional two-power relaxation dependency
like the HN one. Therefore, we consider a more complex
subordination, where a coupling between physical and operational
times ($\gamma$-stable process and its inverse, respectively) is
directed by an independent inverse tempered $\alpha$-stable
process. The coupling leads to two random processes
underestimating and overestimating the real time $t$:
\begin{equation}
T^-_\gamma[S_\gamma(t)]< t < T_\gamma[S_\gamma(t)]\quad {\rm for}\quad t>0\,,\label{eq3}
\end{equation}
where $T_\gamma(\tau)$ is the $\gamma$-stable process, $S_\gamma(t)=\inf\{\tau:
T_\gamma(\tau)>t\}$ is its inverse, and $T^-_\gamma(\tau)=\lim_{x\to\tau_-}T_\gamma(x)$ is
the left-limit $\gamma$-stable process. The random process
$X^-_\gamma(t)=T^-_\gamma[S_\gamma(t)]$ has finite moments of any order whereas $X_\gamma(t)=T_\gamma[S_\gamma(t)]$ has
even no the first moment (see details in \cite{stan10}).

We take into account the process with the finite moments only,
i.e., $X^-_\gamma(t)=T^-_\gamma[S_\gamma(t)]$ with $0<\gamma\leq 1$. Its
probability density is of the form
\begin{equation}
p^-(y,t)=\frac{\sin\pi\gamma}{\pi}\,y^{\gamma-1}(t-y)^{-\gamma}\,,\quad
0<y<t\,.\label{eq4}
\end{equation}
Its moments of any order read
\begin{eqnarray}
\langle X^-_\gamma\rangle&=&\gamma t,\,\langle
(X^-_\gamma)^2\rangle=\frac{\gamma(1+\gamma)}{2}\,t^2\,,\,\dots\,,\,\nonumber\\
\langle
(X^-_\gamma)^n\rangle&=&\frac{\gamma(1+\gamma)\dots(\gamma+n-1)}{n!}\,t^n\,.\nonumber
\end{eqnarray}
Process $X^-_\gamma(t)=T^-_\gamma[S_\gamma(t)]$ is 1-similar and evolves to
infinity as the real time $t$.

To obtain the widely observed fractional two-power relaxation
pattern (\ref{eq0}) we construct a compound subordinator which
combines two following random processes:  one is the coupling
process $X^-_\gamma(t)$, and another is the inverse tempered
$\alpha$-stable process $Q_\alpha(t)$. The process $X^-_\gamma(t)$
directed by $Q_\alpha(t)$ is just the new compound subordinator,
namely $W_{\alpha,\gamma}(t)=X^-_\gamma(Q_\alpha(t))$. We assume
that $X^-_\gamma(t)$ and $Q_\alpha(t)$ are independent on each
other. In this case the anomalous diffusion is determined by the
following relation
\begin{equation}
p^{\,r}(x,t)=\int_0^\infty\int_0^\infty
p^Y(x,y)\,p^-(y,\tau)\,g_\alpha(\tau,t)\,dy\,d\tau\,,\label{eq4a}
\end{equation}
where $p^{\,r}(x,t)$ is the probability density of the
subordinated process obtained from $Y$ directed by
$W_{\alpha,\gamma}(t)$, and $p^Y(x,y)$ is the probability density
of the parent process $Y$. The explicit form of the latter is not
important here. For simplicity, but without loss of generality, as
the parent process can be taken the standard Brownian motion. The
process $X^-_\gamma(t)$ is the essence of time clustering in the
compound subordination. This process is approximated by a simple
continuous-time random walk in which each waiting time is exactly
equal to the jump. Consequently, a walker, moving along a Brownian
trajectory in presence of the subordination by $X^-_\gamma(t)$
(without any government of $Q_\alpha(t)$), stops from time-to-time
and overjumps through intermediate positions in the Brownian
trajectory. In another words the process $X^-_\gamma(t)$ leads to a
random partition (or clustering) of walker's trajectories.

\section{Derivation of relaxation function}
The primary, commonly accepted, interpretation of relaxation
phenomena is based on the concept of exponentially relaxing
objects (for example, dipoles) with different relaxation rates
\cite{J1}. If the relaxing objects do not interact with each
other, then their macroscopic relaxation function is described by
the D law $\phi(t)=\exp(-\omega_pt)$. But the simplest situation,
as it is known from many experiments, is realized very rarely.
Since the relaxing objects can interact with their environment,
their evolution to an equilibrium state has a complex (random
walk-like) behavior which, in general, does not yield the D law.
This interaction may be taken into account with an aid of the
temporal subordination \cite{stan08,stan03}. In this case the
relaxation function reads
\begin{displaymath}
\phi(t)=\int_0^\infty e^{-\omega_p\eta}\,p(\eta,t)\,d\eta\,,
\end{displaymath}
where $p(\eta,t)$ is the subordinator pdf. Particularly, when such
a subordinator is the inverse $\alpha$-stable process $S_\alpha(t)$, the
relaxation function takes simply the Mittag-Leffler (or the
corresponding CC) form. Note that the solutions of fractional
differential equations are just expressed in terms of the
Mittag-Leffler function and its various generalizations (see, for
example, the excellent books \cite{kilbas,klages}). They have
extensive modern-day applications in the study of complex systems,
which maintain the long-memory and nonlocal properties of the
corresponding dynamics such as anomalous diffusion \cite{mk2000}
and non-exponential relaxation
\cite{metz95,Novikov,hilfer2002,taras2008}. In particular,
multi-particle systems with impact phenomena, that exhibit
interactions between particles, have fractional dynamics
\cite{machado}. Nevertheless, the fractional differential
equations are only a macroscopic description of the complex system
evolution, and their microscopic dynamics should be found out from
other reasons.

\begin{figure}
\centering
\resizebox{0.75\columnwidth}{!}{%
  \includegraphics{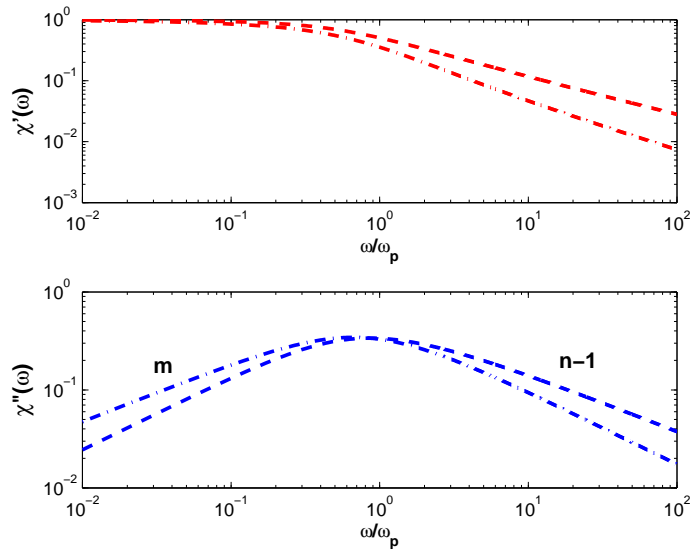}
} \caption{(Color online) Real and imaginary terms of
susceptibility $\chi(\omega)$ (\ref{eq7}), showing the power-law property (\ref{eq0}). Dashed lines represent $m=\gamma=$ 0.75 and $1-n=\alpha=$ 0.6 whereas dash-dotted lines correspond to
$m=\gamma=$ 0.6 and $1-n=\alpha=$ 0.75. The parameter $\sigma$ equals to 0.5.}\label{fig1}
\end{figure}

If the exponentially decayed states are subordinated by the
random process $W_{\alpha,\gamma}(t)$, then the time-domain relaxation
function is expressed in terms of the integral relation
\begin{eqnarray}
\phi(t)=\int_0^\infty\int_0^\infty e^{-\omega_py}\,p^-(y,\tau)
\,g_\alpha(\tau,t)\,dy\,d\tau\nonumber\\
=\frac{\sin\pi\gamma}{\pi}\int^\infty_0\int^1_0e^{-\omega_p\tau
z}\,z^{\gamma-1}\,(1-z)^{-\gamma}\,g_\alpha(\tau,t)\,dz\,d\tau,\label{eq5}
\end{eqnarray}
where $\omega_p$ is a constant. For the experimental study the
frequency-domain representation of the latter function
\begin{equation}
\varphi^*(\omega)=\int^\infty_0e^{-i\omega
t}\,\left(-\frac{d\phi(t)}{dt}\right)\,dt\label{eq6}
\end{equation}
is of more interest. It is well known \cite{J1,J2} that the
complex dielectric susceptibility
$\chi(\omega)=\chi'(\omega)-i\chi''(\omega)\propto\varphi^*(\omega)$
of most relaxing substances shows a peak in the loss component
$\chi''(\omega)$ at a characteristic frequency $\omega_p$. After
integration of Eq.(\ref{eq6}) by means of Eq.(\ref{eq5}) we get
\begin{equation}
\varphi^*(\omega)= 1-\Bigg(\frac{(i\omega/\omega_p+\sigma)^\alpha
-\sigma^\alpha}{1-\sigma^\alpha+(i\omega/\omega_p+\sigma)^\alpha}\Bigg)^\gamma\,,
\label{eq7}
\end{equation}
where $0\leq\sigma=\delta/\omega_p<\infty$ is a positive constant.
Such a frequency-domain relaxation function fits the asymptotic
behavior of the dielectric susceptibility $\chi(\omega)$ with
different low- and high-frequency power tails, namely
\begin{eqnarray}
\Delta\chi(\omega)\propto&\left(i\omega/\omega_p
\right)^\gamma&\qquad{\rm
for}\quad \omega\ll\omega_p\,,\sigma\neq 0,\nonumber\\
\chi(\omega)\propto&\left(i\omega/\omega_p\right)^{-\alpha}&\qquad{\rm
for}\quad \omega\gg\omega_p\,. \label{eq8}
\end{eqnarray}
The main feature of the susceptibility obtained above is that the
indices $m=\gamma$ and $1-n=\alpha$ are determined by different
processes. The index $\alpha$ is connected with the tempered
$\alpha$-stable process only, and $\gamma$ is basically determined
by the coupling process $T^-_\gamma[S_\gamma(t)]$. Thus, by the law (\ref{eq7})
we can describe any fractional two-power relaxation dependency
with both $m>1-n$ and $m<1-n$ (see Fig.~\ref{fig1}). This
advantage allows one to get any point on the relaxation diagram
shown in Fig.~\ref{fig0}, but here is some nuances.

\begin{figure}
\centering
\resizebox{0.75\columnwidth}{!}{%
  \includegraphics{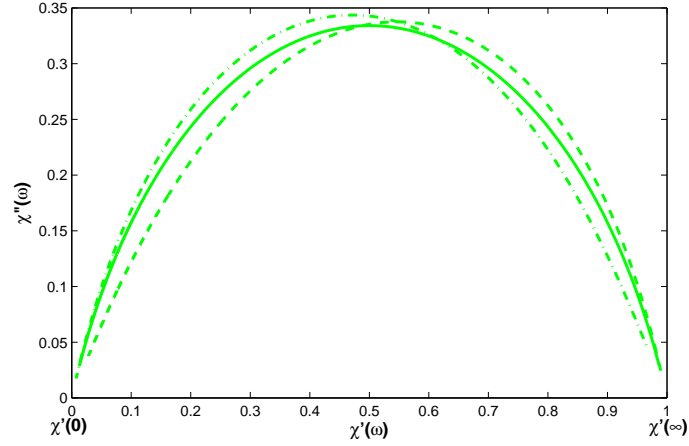}
} \caption{(Color online) Cole-Cole plot for Eq. (\ref{eq7}) 
with the same values of parameters as in Fig.~\ref{fig1}: the solid (symmetric) line
corresponds to the CC susceptibility with $m=1-n=$ 0.75 and $\delta=$ 0, the dashed (asymmetric) line 
relates to the tempered case for $m>1-n$ whereas the dash-dotted line (asymmetric in 
another side) represents the tempered case for $m<1-n$.}\label{fig2}
\end{figure}

Recall that if one uses the conventional $\alpha$-stable process
instead of the tempered one, then in order to come to both
fractional two-power relaxation dependencies, both undershooting $X_\gamma^-(S_\alpha(t))$
and overshooting $X_\gamma(S_\alpha(t))$ processes in the corresponding compound
subordinator should be considered \cite{wjmwt2010}. In particular,
the overshooting process clearly leads to the HN law of
relaxation, but the process itself has no finite moments. Notice
that the ordinary $\alpha$-stable process influences on both
indices ($m$ and $1-n$) in the response function (see
Table~\ref{tab1}), whereas the undershooting process acts only on
the index $m$, and the overshooting process changes only the index
$1-n$. As the indices $0<\alpha,\gamma<1$, the value
$\alpha\gamma<\alpha$ holds true always. With the parameter
$\delta=0$ the tempered $\alpha$-stable process becomes simply
$\alpha$-stable, and the frequency-domain relaxation response
$\varphi^*(\omega)$ coincides with the form arising from the
aforesaid undershooting process in the corresponding compound
subordinator \cite{twsa2011}. In this case the low-frequency
asymptotics $\Delta\chi(\omega)$ is proportional to
$\left(i\omega/\omega_p \right)^{\gamma\alpha}$. The CC law is
just obtained from Eq.(7) for $\sigma=0$ and $\gamma=1$. The CC plot
for Eq. (\ref{eq7}) is shown in Fig.~\ref{fig2}.

\begin{table}
\caption{Asymptotic behavior of the frequency-domain response
function for fractional relaxation scenarios (see also
\cite{wjmwt2010}).} \label{tab1}
\begin{tabular}{p{3cm}p{3.2cm}lc}
\hline\noalign{\smallskip}
Type of relaxation & Operational time (subordinator)& $m$ & $1-n$\\
\noalign{\smallskip}\hline\noalign{\smallskip}
Mittag-Leffler (ML) & $S_\alpha(t)$ & $\alpha$ & $\alpha$ \\
Havriliak-Negami & $X_\gamma(S_\alpha(t))$ & $\alpha$ & $\alpha\gamma$ \\
Generalized Mittag-Leffler (GML) & $X^-_\gamma(S_\alpha(t))$ & $\alpha\gamma$ & $\alpha$ \\
Tempered\\ response (\ref{eq7}) & $X^-_\gamma(Q_\alpha(t))$ & $\gamma$ & $\alpha$ \\
\noalign{\smallskip}\hline
\end{tabular}
\end{table}

When the parameter $\gamma$ is equal to 1, the expression
(\ref{eq7}) simplifies to the form characteristic for the ordinary
tempered relaxation. This is not surprising because in this case
the probability density $p^-(y,\tau)$ becomes the Dirac-delta
function $\delta(\tau-y)$. When the parameter $\alpha$ tends to 1,
the probability density $g_\alpha(t,\tau)$ reads as
$\delta(t-\tau)$. For $\delta=\alpha=1$ we obtain clearly the
ordinary exponential relaxation. Moreover, the real and imaginary
parts of the permittivity data as a function of frequency obtained
for the investigated samples of $\rm Cd_{0.99}Mn_{0.01}Te$:Ga at
various temperatures are well fitted by the dependence (\ref{eq7})
with the parameters $\sigma\approx0$, $\alpha\approx1$ and
$\gamma\approx0.63$ \cite{trz11}. This dependence is also
applicable for appropriate glass-forming materials such as, for
example, ones studied in \cite{J2,jap}.

\section{Conclusions}
This approach gives a chance to include the fractional
two-power-law dependency, widely known from relaxation
experiments, with $m>1-n$ and $m<1-n$ in the framework of tempered
relaxation. Before, with this in mind to describe the diagram of
Fig.~\ref{fig0}, the HN relaxation law was completed for $m < 1-n$
by the GML (Generalized Mittag-Leffler) dependence (see more
details in \cite{wjmwt2010,stanwer10}). Although the new form of
the frequency-domain relaxation function is more complicated than
the HN's law, the latter requires necessarily to account for the
GML relaxation law because a considerable part of experimental
data in relaxation studies cannot be described by any modified
version of the HN law. The proposed attempt also permits one to
select easily effects influencing on the power character of low-
and high-frequency tails in the relaxation function. The
appearance of the passage from $m<1-n$ to $m>1-n$ and vice versa
in the same material reflects structural changes to show what has
a dominant character, either flip-flops of dipoles (relaxing
objects) at short times or cluster patterns of dipole sets at long
times. The short-time behavior can be produced by traps, and the
long-time trend is a result of long-range interactions of dipoles
via their cluster and collective regions. It should be stressed
that in this scheme we use only the random processes with finite
moments what is important for analysis and interpretation of
experimental data \cite{J2}.

\section*{Acknowledgments}

Work of K.W. was partially supported by the Ministry of Sciences
and Higher Education project PB NN 507503539. A.S. is grateful to
the Institute of Physics and the Hugo Steinhaus Center for
pleasant hospitality during his visit in Wroc{\l}aw University of
Technology.


\end{document}